\documentclass[10pt,twocolumn,letterpaper]{article}

\usepackage{iccv}
\usepackage{times}
\usepackage{epsfig}
\usepackage{graphicx}
\usepackage{amsmath}
\usepackage{amssymb}
\usepackage{mathtools}
\usepackage{bbm}
\usepackage{makecell, verbatim, sidecap}
\usepackage{booktabs}
\usepackage{multirow}
\usepackage[breaklinks=true,bookmarks=false]{hyperref}

\iccvfinalcopy %

\usepackage{color}

\def\block1{{$\sf block1$}}
\def\block2{{$\sf block2$}}
\def\block3{{$\sf block3$}}
\def\block4{{$\sf block4$}}

\usepackage[margin=4pt,font=small,labelfont=bf,labelsep=endash,tableposition=top]{caption}

\renewcommand{\texttt}[1]{ $ {{\tt #1} } $}

\usepackage{enumerate}
\usepackage{xspace}
\usepackage{float}  
\usepackage{subfigure}  

\usepackage[T1]{fontenc}

\renewcommand{\vec}[1]{\ensuremath{\pmb{#1}}}
\newcommand{\mat}[1]{\ensuremath{\mathbf{#1}}}
\newcommand{\set}[1]{\ensuremath{\mathscr{#1}}}

\makeatletter
\@tfor\next:=abcdefghijklmnopqrstuvwxyxz\do
 {\begingroup\edef\x{\endgroup
    \noexpand\@namedef{v\next}{\noexpand\vec{\next}}%
  }\x}
\@tfor\next:=ABCDEFGHIJKLMNOPQRSTUVWXYZ\do
 {\begingroup\edef\x{\endgroup
    \noexpand\@namedef{m\next}{\noexpand\mat{\next}}%
  }\x}
\@tfor\next:=ABCDEFGHIJKLMNOPQRSTUVWXYZ\do
 {\begingroup\edef\x{\endgroup
    \noexpand\@namedef{s\next}{\noexpand\set{\next}}%
  }\x}
\makeatother

\def\Ours{{FCNDet}\xspace}
\def\Ours{{DPN-SENet}\xspace}

\begin{document}

\title{\textbf{\Ours:A self-attention mechanism neural network for 
	detection and diagnosis of COVID-19 from chest x-ray images}}

\author{
Bo Cheng ~ ~ ~ ~ ~ 
Ruhui Xue ~ ~ ~ ~ ~ 
Hang Yang ~ ~ ~ ~ ~ 
Laili Zhu ~ ~ ~ ~ ~
Wei Xiang\thanks{Corresponding author,email: \texttt{21500068@swun.edu.cn}
}
\\
Southwest Minzu University, China 
}

\maketitle
\thispagestyle{empty}

\begin{abstract}
\textbf{Background and Objective}: The new type of coronavirus is also called COVID-19. 
It began to spread at the end of 2019 and has now spread across the world. Until October 
2020, It has infected around 37 million people and claimed about 1 million lives. We 
propose a deep learning model that can help radiologists and clinicians use chest X-rays to diagnose COVID-19 cases and show the diagnostic features of pneumonia.
 
\textbf{Methods}: The approach in this study is: 1) We propose a data enhancement method 
to increase the diversity of the data set, thereby improving the generalization 
performance of the model. 2) Add a self-attention mechanism to the DPN network. The addition of a self-attention 
mechanism has greatly improved the performance of the network. 3) Use the Lime
interpretable library to mark the feature regions on the X-ray medical image that are 
help doctors more quickly diagnose COVID-19 in people.

\textbf{Results}: Under the same network model, the data with and without data enhancement 
is put into the model for training respectively. At last, comparing two experimental 
result: among the 10 network models with different structures, 7 network models have 
improved their effects after using data enhancement, with an average improvement of 
1\% in recognition accuracy. We propose that the accuracy and recall rates of the DPN-SE network are 93\% and 98\% of cases (COVID-19 vs. pneumonia bacteria vs. viral pneumonia vs. normal). Compared with the original DPN, the respectively accuracy is improved by 2\%. In terms of picture interpretability, our interpretability model can mark the key feature areas of the X-ray image.

\textbf{Conclusion}: The data augmentation method we used has achieved effective results on a small amount of data set, showing that a reasonable data augmentation method can improve the recognition accuracy without changing the sample size and model structure. Our proposed network model DPN-SE with increased attention mechanism structure can improve the recognition accuracy on the original basis. Overall, the proposed method and model can effectively become a very useful tool for clinical radiologists.

\end{abstract}

\section{Introduction}
At the end of 2019, a series of new type pneumonia cases was identified, which was 
caused by the novel coronavirus that was first discovered in 2019.This epidemic was 
declared constitutes a Public Health Emergency of International Concern in on January 
2020 and it was officially named as COVID-19 on February 2020 by WHO \cite{Wu2020}. Then 
COVID-19 began to spread worldwide. Until October 2020, there are 37 million 
confirmed cases and 1 million people have lost their lives.

Coronaviruses are a group of viruses that cause respiratory tract infections. These 
viruses are very common in animals, but in some cases it can mutate to infect humans, 
and then spread rapidly among people. SARS-CoV-2(The virus cause the COVID-19\cite{Yi2020}) 
is the seventh known coronavirus that can infect humans. SARS-CoV-2 can cause 
respiratory tract infections, Intestinal infections and fewer neurological symptoms. 
Some infected person of SARS-CoV-2 may faces various sequela after recovery.

After the outbreak of the epidemic, the confirmed cases increased exponentially. As 
of 6:30 am September 22nd in Beijing time, there are 31,452,367 confirmed cases 
reported around the world, tragically 968,239 people around the world have lost their 
lives to this virus and 96 countries confirmed more than 10,000 cases. In the Americas, 
U.S. Covid-19 death toll approach 200,000. In Africa, the epidemic in Morocco 
rebounded after loosening control measures.At present, the epidemic is still spreading 
all over the world, and the risk of sporadic cases and local outbreaks in China still exists. 
On September 18, 2020, at the Zhongguancun Forum Global Science and life healthforum, Zhong Nanshan, academician of the Chinese Academy of engineering, revealed 
that the epidemic may still exists in this winter and next spring. It also means that 
COVID-19 and respiratory diseases such as influenza will come together with the 
advent of autumn and winter. The task of prevention and control epidemic will be very 
difficult.

Current evidence suggests that the primary way the virus spreads is by respiratory 
droplets\cite{pub.1134389844,Fanelli2020} among people who are in close contact with each other. The incubation 
period of COVID-19 is generally 1 to 14 days, and the longest is 24 days. In order to 
effectively prevent and control the epidemic, we not only need to wear masks, wash 
hands frequently, measure body temperature every day, but also need to trace the source 
of the virus.

In the early stage of the outbreak, due to the explosive growth of the number of
confirmed cases, the demand for suspected case diagnosing also increased, which put 
great pressure on the medical system with scarce medical resources. The COVID-19 
detection reagents currently approved for use mainly include two types, one is nucleic 
acid detection reagent, and the other is antibody detection reagent. Nucleic acid 
detection is the main detection method at present, but it takes at least a few hours and a 
maximum of several days, and it has a high professional threshold for operators. And a 
more fatal drawback is the high probability of false negatives using nucleic acid testing. 
The advantage of antibody detection is that it takes a short time and the risk to medical 
staff is lower. However, in the early stage of virus infection, antibody may not be 
produced in human body, which makes the detection window period exist and there is 
the possibility of missing detection. Therefore, antibody detection is generally used for 
auxiliary detection of negative cases that test by nucleic acid or rapidly screening 
potential cases in the population, but it cannot replace nucleic acid detection.

The limitations of detection methods and the current global epidemic situation 
encourage us to try to propose a deep network learning model \cite{Fanelli2020,Wang_2020,Gozes2020}. We use self-attention mechanism and advanced image classification algorithm to intelligently 
diagnose and quantitatively evaluate X-ray images, and marked the severity of various 
pneumonia diseases including local lesions, diffuse lesions and whole lung involvement,which can help doctors identify lesions more quickly and accurately. The model focus 
on the lesion site, doctors can directly focus on the labeled part and make diagnosis 
quickly, which greatly shortens the diagnosis time of COVID-19.

\begin{figure*}[ht!]
	\begin{center}
		\includegraphics[width=.85\linewidth]{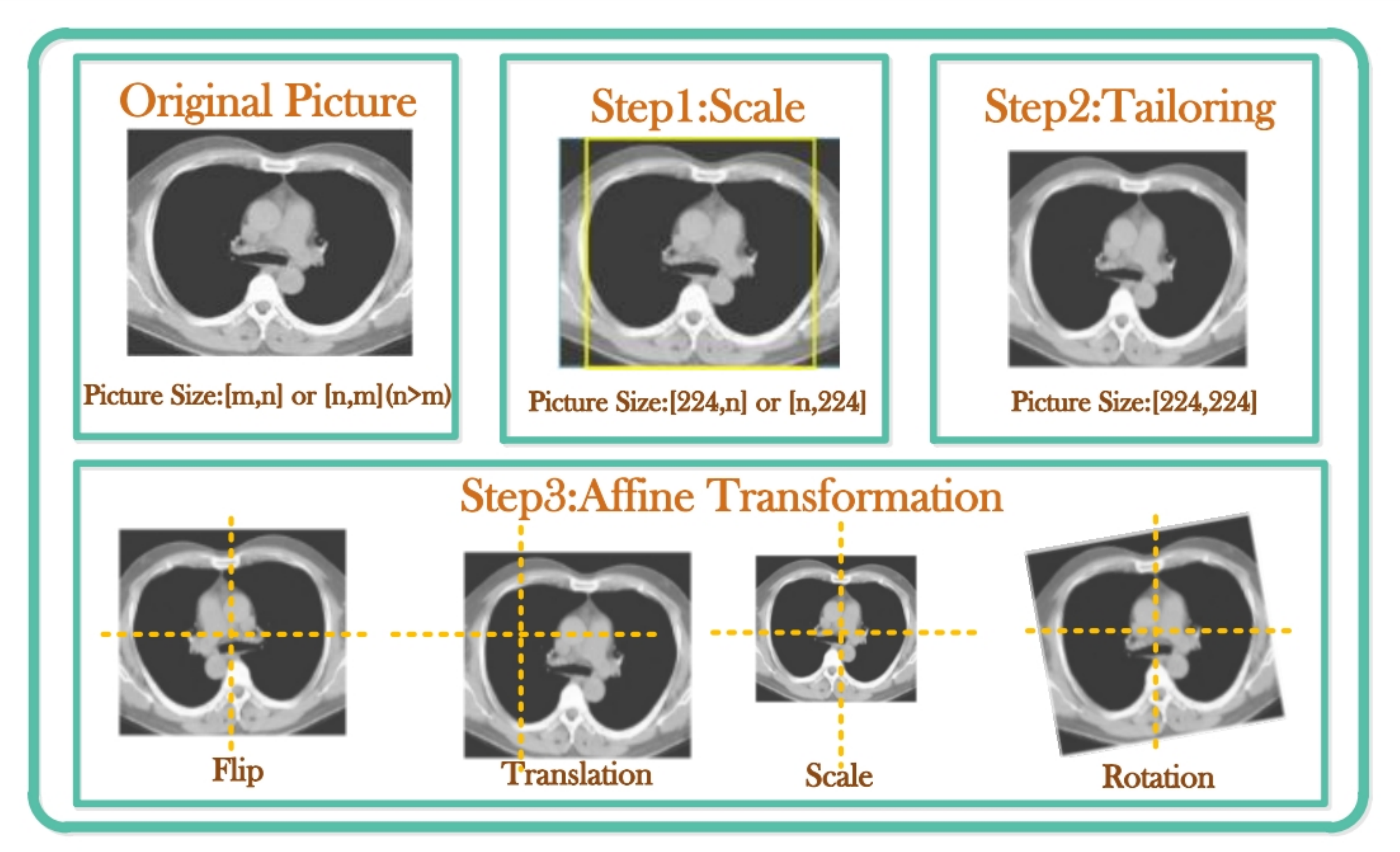}
	\end{center}
	\caption{ Overview of the Data Augmentation. Step1: Scale the narrow side of the image to size 224; 
		Step2: Randomly tailoring a picture of size [224, 224] from an image of [224, n] or [n,224]; Setp3: The 
		[224, 224] size picture is processed by affine transformation: flip, rotate, scale.
	}
	\label{fig:data_enhancement _process}
\end{figure*}

\section{Related Work}
With the rise of artificial intelligence, deep learning has been gradually used in 
health-related fields, and its wide use in other fields has proved that deep learning can 
help people solve some basic problems. In this chapter, we review and summarize the 
literature on the application of artificial intelligence technology to detect chest diseases

With the global outbreak of covid-19, people in different fields have made different 
contributions to against this epidemic. The pioneers of artificial intelligence through
the technology of image classification to classify chest CT and X-ray images\cite{Hall2020,Alimadadi2020,Chen2020,Gore_2020,Hemdan2020,2020Corona}
. Each other put forward different deep learning network architecture to diagnose patients.
With the rapid development of artificial intelligence, computer classification of images 
has been proven to have a higher accuracy rate than human eye recognition \cite{Apostolopoulos_2020} (many 
network models have better classification effects on ImageNet data sets than ordinary 
people's judgment effects). Many researchers are committed to improving the detection 
and analysis methods of various diseases by acquiring radiology data sets and applying 
data science classifiers

Hemdan \cite{Narin_2021} used a deep learning model to diagnose COVID-19 in X-ray images 
and proposed a COVIDX-Net model with 7 convolutional layers. Wang and Wong \cite{Wang2020}
proposed a deep model (COVID-Net) for COVID19 detection, which achieved 92.4\% 
accuracy when classifying normal, non-COVID pneumonia and COVID-19. Ioannis 
developed a deep learning model using 224 COVID-19 images \cite{Song_2021}. Their model 
achieved accuracy rates of 98.75\% and 93.48\% in 2 and 3 categories, respectively. 
Narin used chest X-ray images coupled with the ResNet50 model to obtain 98\% 
COVID-19 detection accuracy \cite{Wang_2021}. Sethy and Behera used a support vector machine 
(SVM) classifier to classify features obtained from various convolutional neural 
network (CNN) models using X-ray images \cite{Sethy_2020}. It shows that the ResNet50 model
with SVM classifier has the best performance. Finally, there are some recent studies on 
COVID-19 detection. These studies use various deep learning models with CT images 
\cite{thno45016,Das_2020,Zheng_2020,Xu2020,barstugan2020coronavirus,chen2020residual}.

Li designed the COVNet neural network for COVID-19 detection to extract visual 
features from CT images \cite{Li_2020}. Their model uses a RestNet50 network with input CT 
images. They performed 4,356 chest CT scans on 3,322 patients (1296 CT scans for 
COVID-19, 1,735 CAP and 1,325 non-pneumonia), with an accuracy equal of 0.96.

Fan et al.  proposed an Inf-Net\cite{Fan2020} segmentation model to automatically identify infected areas from CT slices. In Inf-Net, a parallel partial decoder is used to aggregate the high-level features and generate a global map. Then, the implicit reverse attention and explicit edge-attention are utilized to model the boundaries and enhance the representations. The first model is a network based on 
ResNet18, then the second model is designed on the first network to construct a 
structure by connecting the location attention mechanism in the fully connected layer, 
with the purpose of improving the overall accuracy. Their dataset consists of 1,710 CT 
images, including 357 COVID-19, 390 influenza A viral pneumonia, and 963 normal.

\section{Method}
In this section, we will discuss the data processing methods, the implementation of 
model architecture and training methods.

\subsection{Data Augmentation}
Because it is difficult to obtain chest X-ray images of patients with new coronary 
pneumonia, we have collected fewer chest X-ray images of patients with new coronary 
pneumonia (the training set contains 1119 images, and the test set contains 293 images, 
for a total of 1412 images). Using a small number of data to train the network model 
can easily cause the model to overfit, resulting in very low recognition accuracy on the 
test set. In the field of computer vision, image enhancement is a common implicit 
regularization technique to reduce overfitting in deep convolutional neural networks, 
and is widely used to improve the performance of benchmark datasets.\cite{ratner2017learning,cubuk2019autoaugment}.

Common image enhancement methods include certain variations and combinations of flipping, rotating, scaling, and cropping. Different fields, imaging methods and tasks may benefit from a variety of image transformations and combinations \cite{ratner2017learning}. For 
example, in the medical image analysis done in this article, compared with natural 
images, the data set is usually small and difficult to obtain, and the details in the image 
are very important, which may be the basis for doctors to distinguish diseases. So when 
we do data enhancement, we should not modify the original image too much, otherwise 
the details on the x-ray image will be eliminated. After many attempts in the 
experiments of this paper, it is found that the data enhancement process shown in Fig.~\ref{fig:data_enhancement _process}  can achieve better results. The whole process goes through three steps, step1: Scale 
the narrow side of the image to size 224; setp2: Randomly tailoring a picture of size 
[224, 224] from an image of [224, n] or [n, 224] size; setp3: The [224, 224] size picture 
is processed by affine transformation: flip, rotate, scale.
\begin{figure*}[ht!]
	\begin{center}
		\includegraphics[width=.85\linewidth]{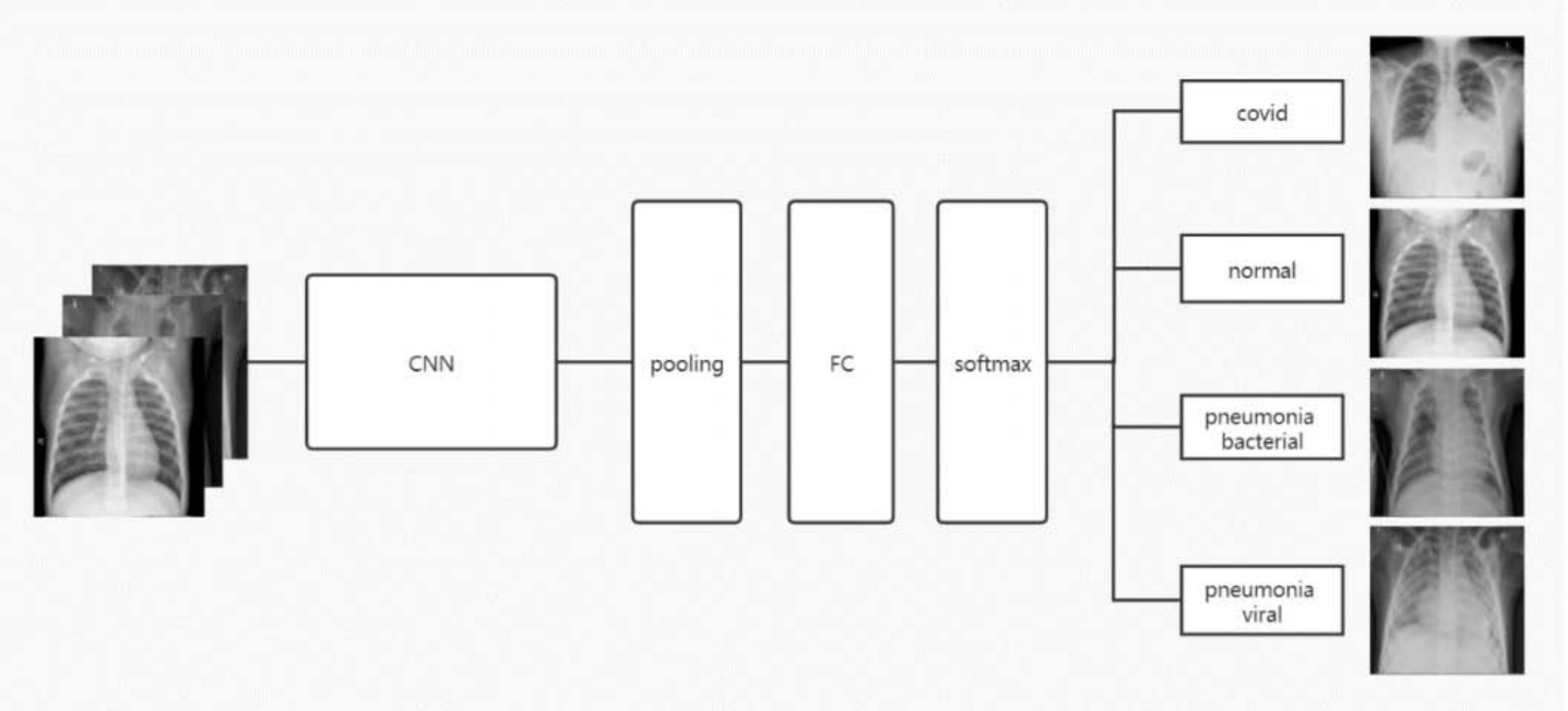}
	\end{center}
	\caption{Network overall framework diagram
	}
	\label{fig:Network overall framework diagram}
\end{figure*}
\subsection{Network Model}
In this section, we will discuss the whole network framework, which is based on the convolutional neural network in deep learning to classify covid19 x-ray images. The 
overall framework is shown in Fig.~\ref{fig:Network overall framework diagram}. The main framework of the network is the 
DPNNet \cite{chen2017dual}, and the structure is shown in Figure 1. This network combines the 
advantages of ResNet \cite{He2015} and DenseNet \cite{huang2018densely}, and won the championship in the 2017 
ImageNet classification competition \cite{Deng2009}. In this paper, we use a 92-layer DPN network, 
which started with a 7*7 convolutional layer and maximum pooling layer. Then there 
are 4 stages, each stage is composed of multiple substages. The core idea of substage is 
to integrate ResNet and DenseNet. ResNet: focuses on the reuse of features and the 
sharing of parameters; DenseNet: focuses on the discovery of new features and Achieve 
complementary advantages and disadvantages. Followed by an average pooling layer 
and a fully connected layer, and finally a softmax layer. However, because DPN-92 is 
composed of a large number of convolutional layers, and the convolution kernel, as the 
core of the convolutional neural network, generally can only obtain feature maps from 
the local receptive field, which lacks the characteristics of the global receptive field. So
we introduced the attention mechanism network on the basis of DPN-92.

The human attention visual mechanism uses limited attention to quickly screen high-value information from a large amount of information. The attention mechanism in deep 
learning is essentially similar to the selective visual attention mechanism of humans. 
Its core goal is to select information that is more critical to the current task goal from a 
large number of information. At present, the attention mechanism network has been 
widely used in various deep learning applications such as natural language processing,
image recognition, and speech recognition. In this article, we use the senet attention 
mechanism network.

Senet is mainly composed of three parts: compression part, excitation part and 
reweight part \cite{hu2019squeezeandexcitation}. The compression part compresses the features along the spatial 
dimension, turning each two-dimensional feature channel into a real number. This real 
number has a global receptive field to some extent, and matches the dimension of the 
input and the number of feature channels of the input. It represents the global 
distribution of the response on the feature channel. This part is implemented using 
global pooling.

\begin{figure*}[ht!]
	\begin{center}
		\includegraphics[width=.85\linewidth]{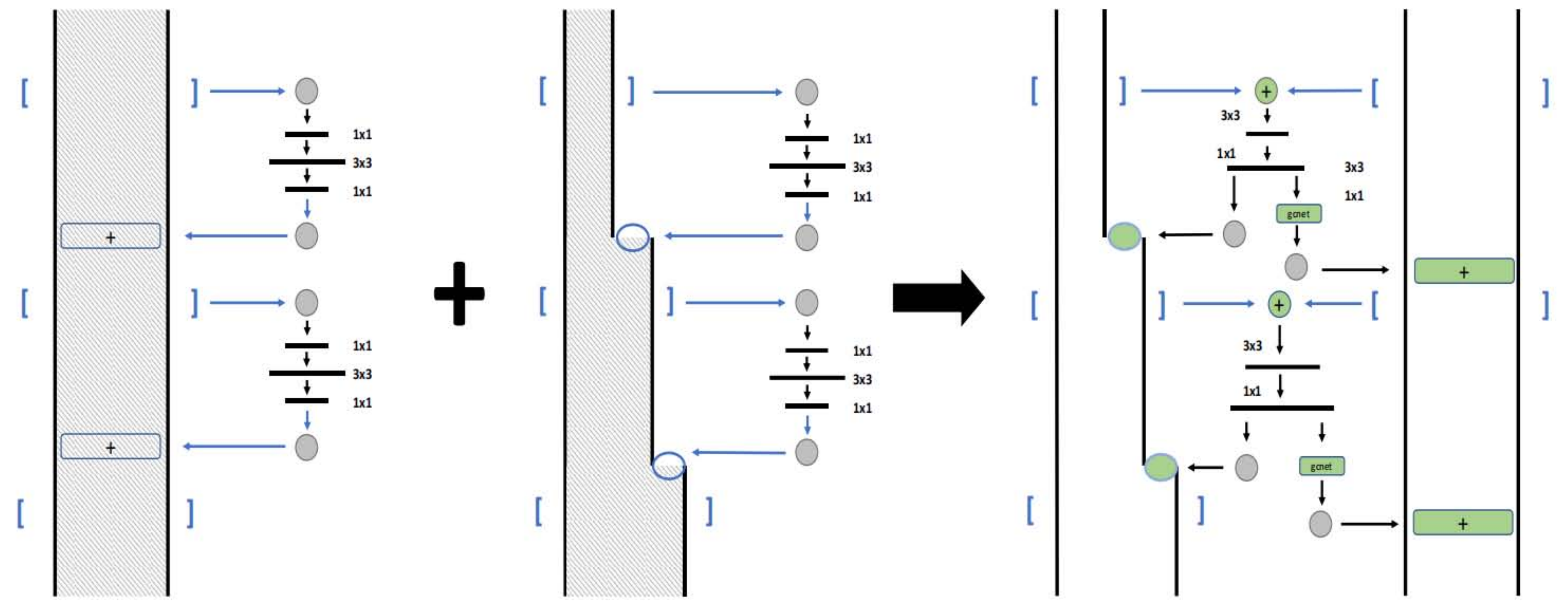}
	\end{center}
	\caption{Network overall framework diagram
	}
	\label{fig:Framework of DPN-SE}
\end{figure*}

The incentive part is similar to the gate mechanism in a recurrent neural network \cite{Fanelli2020}. 
A weight is generated for each feature channel through parameters, and a correlation 
between feature channels is modeled by learning parameters. The last part is reweight,
and we regard the weight of the output of the excitation part as the weight of each 
feature after feature selection The degree of importance is then weighted channel by 
channel by multiplication to the first input feature to complete the recalibration of the 
input feature in the channel dimension. The senet network is introduced into the DPN-
92 network, and the network part becomes the DPN- SE network, as shown in Fig.~\ref{fig:Framework of DPN-SE}.
\subsection{Model Explanation}
With the rapid development of artificial intelligence, artificial intelligence has also 
had a profound impact on all walks of life. From Meitu software to facial recognition 
in train stations, systems built around artificial intelligence have had a huge impact on 
medical care, transportation, criminal justice, financial risk management, and other 
areas of society. Although artificial intelligence is widely used, machine learning 
models are still black boxes. Although these models show capabilities beyond humans, 
we are not yet sure what specific information in the input data will make them make 
decisions. Deep learning models usually contain deeply nested non-linear structures, 
making them less transparent. The lack of transparency is not a problem for AlphaGo 
itself. However, in areas where interpretability and transparency are critical (such as 
medical diagnosis, military and combat operations), the opacity of the model greatly 
hinders the expansion of AI/ML.

From a broad perspective, interpretability \cite{zeiler2013visualizing,doshivelez2017rigorous} means that when we need to solve 
a thing or make a decision about it, we need to obtain information from this thing that 
can be understood enough to help us make a decision. Therefore, in this article, we need 
to clearly mark the basis for the model judgment on the chest X-ray image, so that the 
doctor can quickly find the lesion area and judge whether the marked part is the basis 
for judgment.

In the study of local interpretability of machine learning models, a representative 
method is Local Interpretable Model-Agnostic Explanation (LIME) \cite{Ribeiro2016} proposed by 
Marco Tulio Ribeiro et al. The main idea of LIME is to use interpretability models (such 
as linear models, decision trees) to locally approximate the prediction of the target black 
box model. This method does not go deep into the model. By slightly perturbing the 
input, detecting changes in the output of the black box model, and training an 
interpretable model at points of interest (original input) based on this change. It is worth 
noting that the interpretability model is a local approximation of the black box model, 
rather than a global approximation, which is the origin of its name. The mathematical 
expression of LIME is as follows:$$explanation(x) = \underset{g\epsilon G}{arg min} \  L(f,g,\pi x) + \Omega(g) $$

For the interpretation model $g$, we compare the approximation between  $g$ and the 
original model $f$ by minimizing the loss function, where $ \Omega(g) $ represents the model 
complexity of the explanation model $g$,and $G$ represents all possible explanation 
models(If we use linear models to explain, $G$ means all linear models),$\pi x$ defines the 
realm of $x$.We make the model interpretable by minimizing $L$. Among them, the 
domain size and complexity of the model need to be defined.

Below is a brief description of LIME's workflow. In order to be independent of 
the model, LIME will not go deep into the model. In order to find out which part of 
the input contributes to the prediction result, we will make a slight disturbance 
around the input value and observe the prediction behavior of the model. Then we 
assign weights based on the distance of these disturbed data points from the 
original data, and learn an interpretable model and prediction results based on 
them. The demonstration diagram of this process is as follows. The decision 
function of the original model is represented by a blue/pink background, which is 
obviously non-linear. The bright red cross indicates the sample being interpreted 
(called X). We sample around X and assign weights according to their distance to 
X (where weight means size) We use the original model to predict these disturbed 
samples, and then learn a linear model (dotted line) to approximate the model 
well near X. Note that this explanation only holds near X and is invalid for the 
whole world.

\begin{figure}[t!]
	\centering
	\vspace{-1cm} 
	\subfigtopskip=-2pt 
	\subfigbottomskip=-1pt 
	\subfigcapskip=-8pt 
	\subfigure[]{
		\label{structure (a)}
		\includegraphics[width=\linewidth]{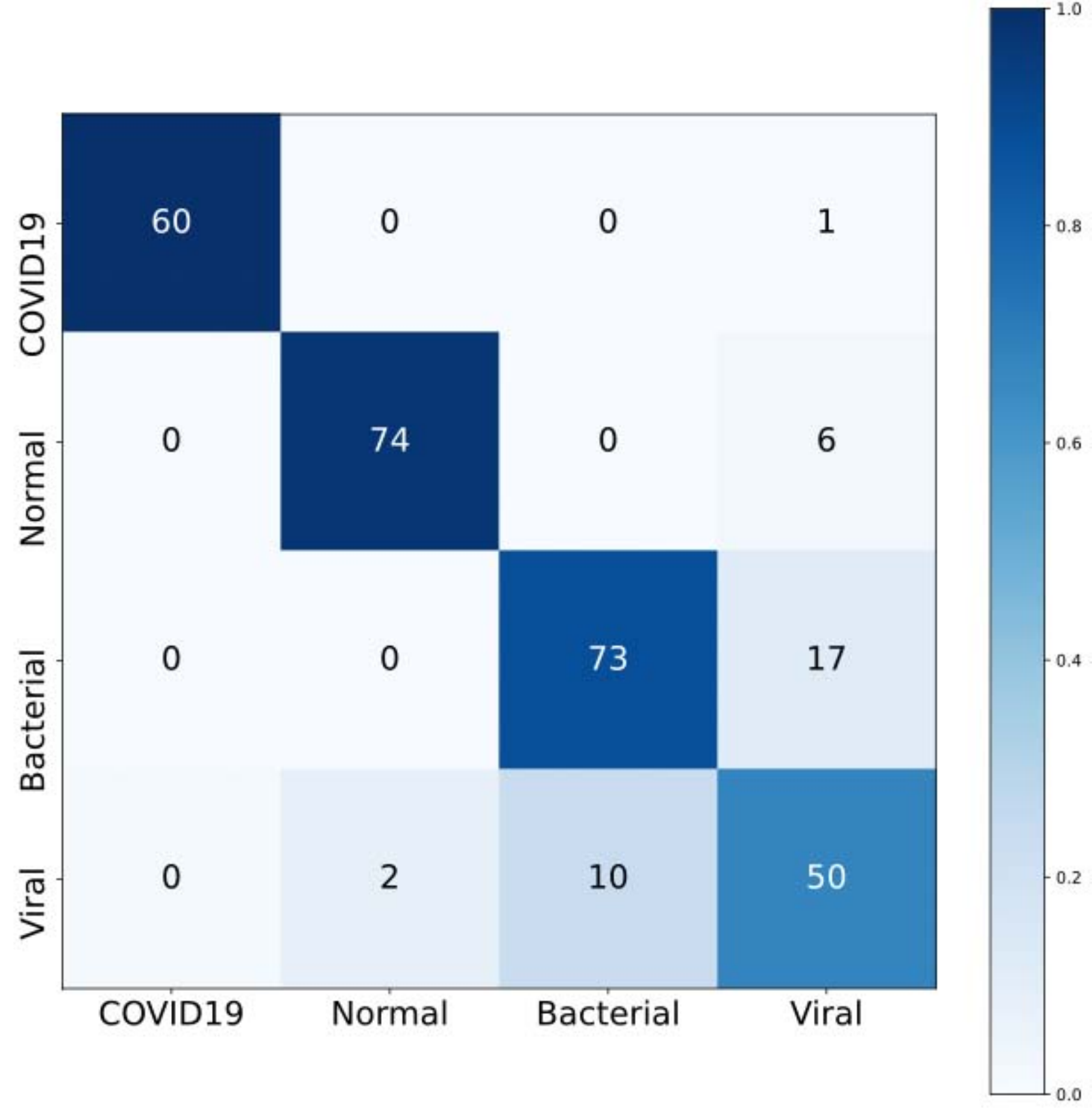}}
	
	\subfigure[]{
		\label{structure (b)}
		\includegraphics[width=\linewidth]{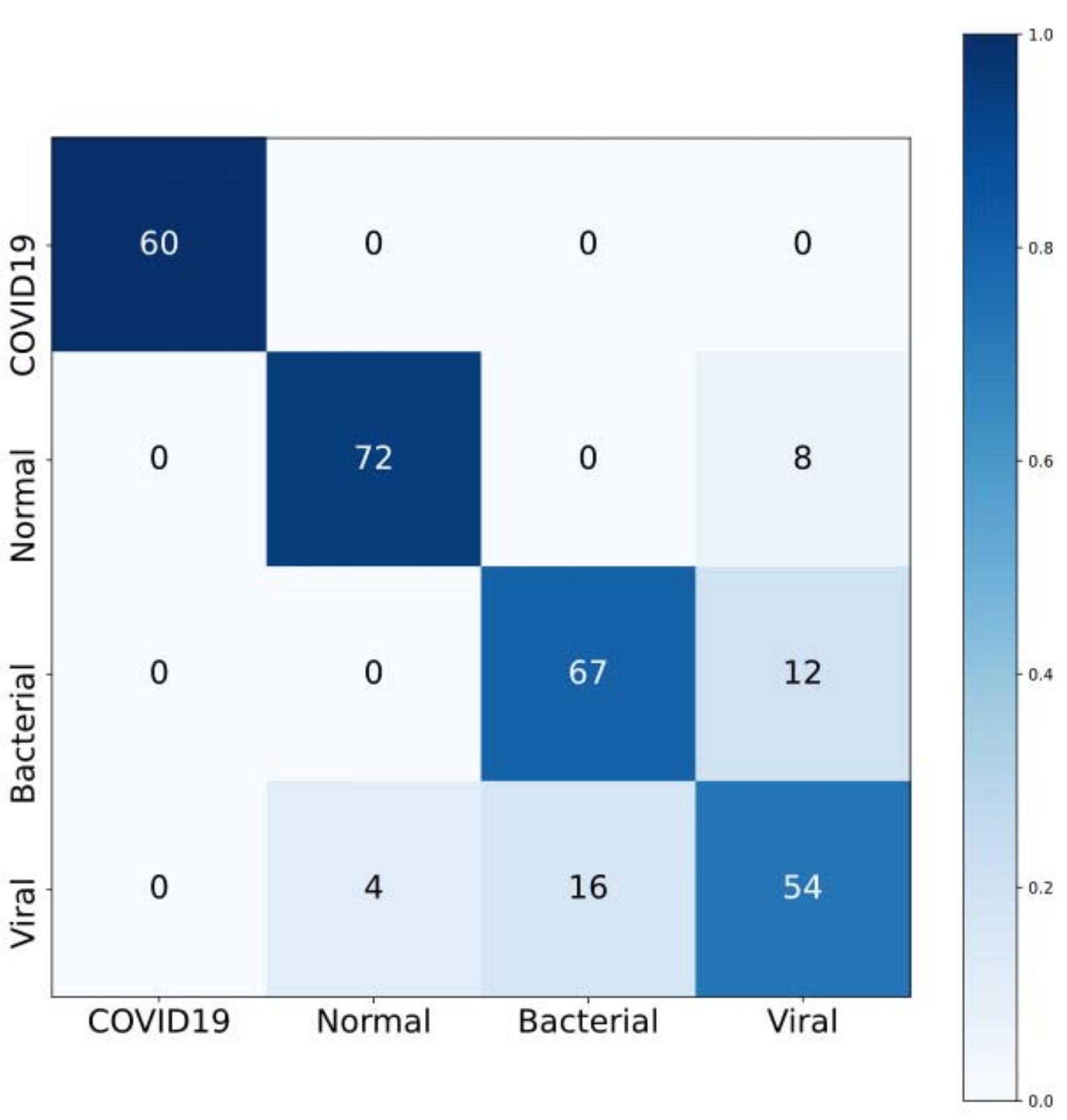}}
	\label{fig:training_targets}
	\caption{Model trained using resnet52 network structure (a) without using data augmentation method(b) using data augmentation method}
	\label{the_fig4}
\end{figure}

\section{Experiment}
In this section, we will introduce the basic information of the data set and analyze 
the experimental data results according to the following aspects:
\begin{enumerate}[(1)]
	\item Detect the effects of using data enhancement and unused data enhancement;
	\item Use the DPN-SE network model to test the recognition accuracy;
	\item Use an interpretable model to highlight the key areas of the X-ray picture of the 
	new crown.
\end{enumerate}

\subsection{Dataset}
The accuracy of model training largely depends on the data set. COVID-19 is a new 
disease. We need to select a large number of chest X-rays to allow our model to fully 
learn the characteristics of the lungs. The X-ray images of COVID-19 are available on 
GitHub by Joseph et al. \cite{cohen2020covid19}.

\begin{figure*}[h]
	\centering  
	\vspace{-0.35cm} 
	\subfigtopskip=1pt 
	\subfigbottomskip=2pt 
	\subfigcapskip=-1pt 
	\subfigure[Covid-19]{
		\label{X-ray images Covid-19}
		\includegraphics[width=0.4\linewidth,height=5cm]{latex/figures/fig4_a.pdf}}
	\quad 
	\quad 
	\quad 
	\subfigure[Normal]{
		\label{X-ray images Normal}
		\includegraphics[width=0.4\linewidth,height=5cm]{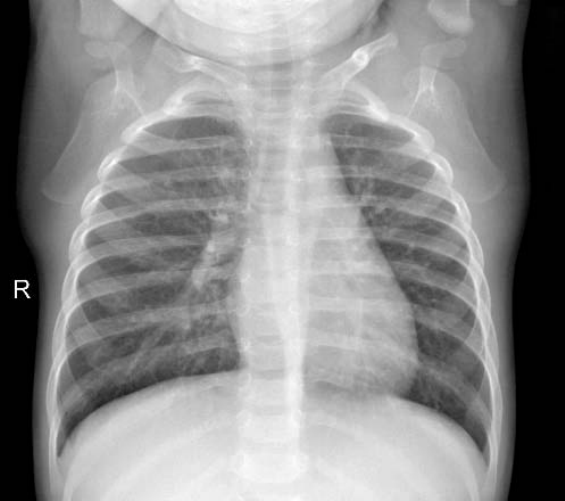}}
	\subfigure[Pneumonia Bacterial]{
		\label{X-ray images Pneumonia Bacterial}
		\includegraphics[width=0.4\linewidth,height=5cm]{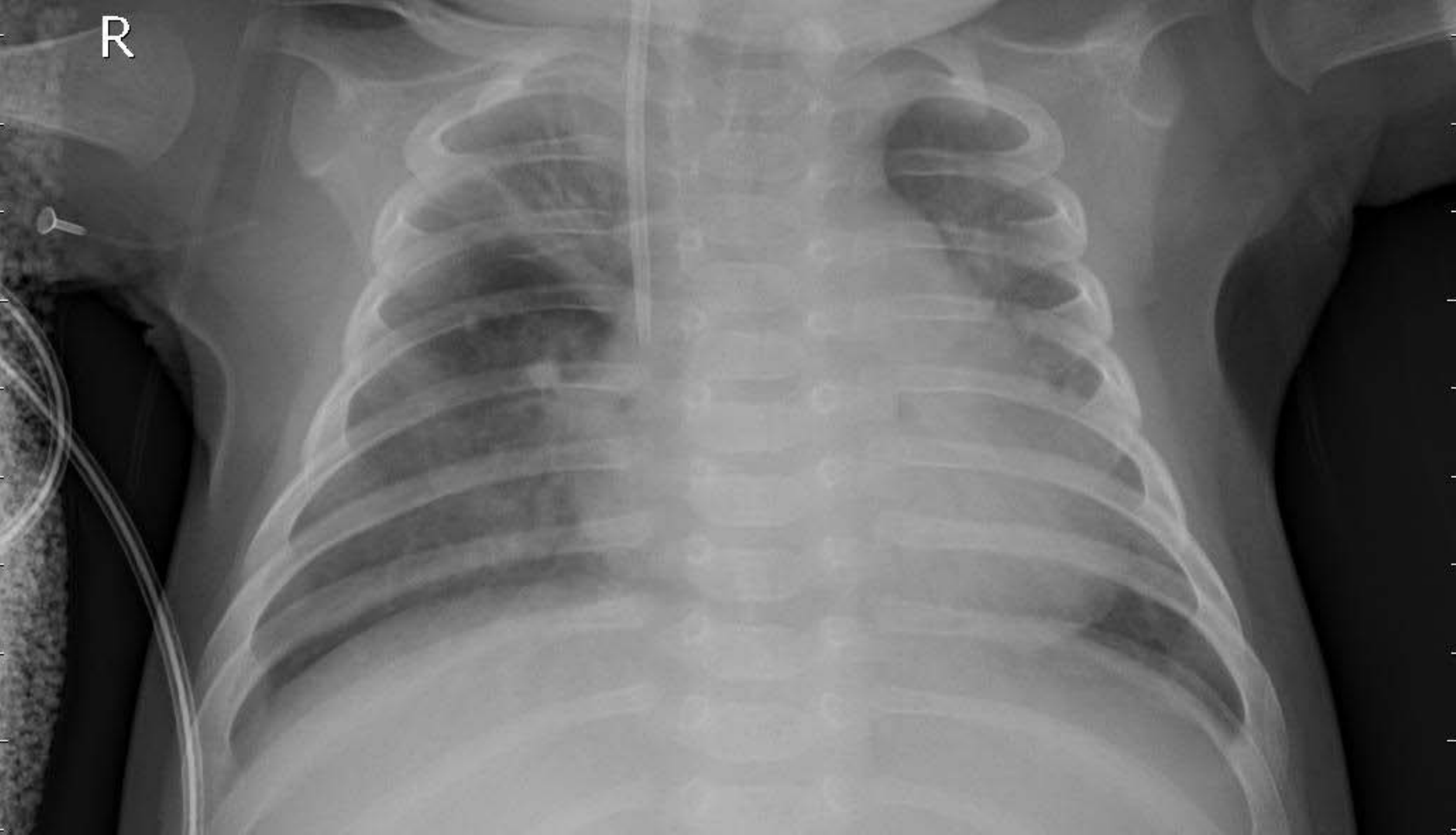}}
	\quad
	\quad 
	\quad 
	\subfigure[Pneumonia Viral]{
		\label{X-ray images Pneumonia Viral}
		\includegraphics[width=0.4\linewidth,height=5cm]{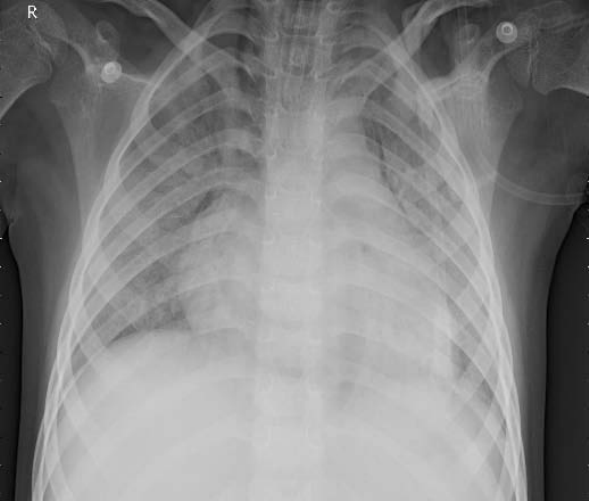}}
	\caption{Types of lung X-ray images}
	\label{Types of lung X-ray images}
\end{figure*}

The author collected chest X-ray images of new coronary pneumonia from various 
real sources of the Radiological Society of North America. Our data set is a four-category data set, which not only contains chest X-ray radiographic images of new 
coronary pneumonia, but also contains bacterial pneumonia, viral pneumonia and 
normal pneumonia from the Kaggle repository "Chest X-Ray Images (Pneumonia) \cite{Kermany2018}
Chest X-ray image. This data set consists of 1203 normal chest radiographs, 660 
bacterial pneumonia chest radiographs and 931 viral pneumonia chest radiographs. The 
data set for our model training comes from the data set compiled by \cite{KHAN2020105581}. Among them, 
the new crown Pneumonia pictures: 304 pictures, normal pictures: 375 pictures, 
bacterial pneumonia: 379 pictures, viral pneumonia: 354 pictures. In order to avoid 
over-fitting problems in the later stage, we used data enhancement technology. The four 
categories of pictures are shown in Fig.~\ref{Types of lung X-ray images}.
\subsection{Results and Discussion}
In this study, the chest X-ray dataset mentioned in Section 4.1 was used to train the 
model. The network structure we use is VGG16, ResNet, InceptionV4, DenseNet, 
DPNNet and our own DPN-SE network with channel self-attention mechanism added. 
First, compare the trained network after data enhancement with the directly trained 
network (the method of data enhancement is introduced in section 3.1). In the 
experiment, all the network models were trained for 100 epochs, and the cost loss graph 
was observed, and finally reached the convergence state. The four performance 
indicators for evaluating the classification model are:$$Accuracy = \frac{TN + TP}{TN + TP + FN + FP} \eqno(1) $$ $$Recall=\frac{TP}{TP + FN} \eqno(2) $$ $$Precision=\frac{TP}{TP + FP} \eqno(3) $$ $$F1-Score=2 \times \frac{Precision \times Recall}{Precision + Recall} \eqno(4) $$

TN, TP, TN, and FN in formulas (1)-(4) represent the number of true, false positive, 
true negative, and false negative, respectively. TP is the proportion of model prediction 
results that are correctly marked as positive; FP is the proportion of incorrectly marked 
as positive; TN is the proportion of correctly marked as false, and FN is the proportion 
of incorrectly marked as false (COVID- 19 is the positive category, and the other categories are the negative category).

In Fig.~\ref{the_fig4}., we present the recognition results trained by the DPN131 network structure 
in the form of a confusion matrix (CM), a) the image is the recognition effect without 
image enhancement, and b) the image is the recognition effect with image enhancement. 
In the comparison of confusion matrix a) and b), we can observe that after adding data 
enhancement, the average accuracy rate has risen from 0.8089 to 0.8328. The accuracy 
rate for the COVID-19 category rose from 0.92 to 0.97, the recall rate rose from 97%
to 98, and the F1-score rose from 94\% to 98\%. Several other categories of indicators 
have also increased. In order to add more verification examples, we used other models 
for comparison. As shown in Table.1, a variety of different network models are used to 
test the effects of data enhancement and unused data enhancement. The evaluation 
indicators include accuracy, precision, recall, F value The calculation of the indicator counts the COVID-19 category as a positive 
example, and the others as a negative example).

As can be seen from the data in Table.~\ref{table1}, we have used 10 network models such as 
ResNet, DenseNet, DPNNet, VGG16, Inceptionv4, etc. to test. The recognition 
accuracy of most models is above 80\%. Comparison of network models using ResNet50. 
If the data set without data enhancement is used to train the network, the average 
accuracy obtained is 79\%, the precision is 0.92, the recall rate is 0.98, and the F-measure is 0.95. If the network is trained on the data set processed by data enhancement, the 
average accuracy obtained is 80\%, the precision is 0.97, the recall rate is 0.98, and the 
F-measure is 0.98. It can be seen from the comparison that after data enhancement
processing, the trained network model can improve the recognition accuracy by 1\%. 
Among the 10 network models displayed in total, the recognition effect of 7 network 
models has improved after data enhancement processing, the recognition effect of 2 
network models has decreased, and the recognition effect of 1 network model has 
basically remained unchanged. Finally, we can infer that after processing the x-ray 
sample data using data enhancement methods, training the network model can 
effectively improve the recognition accuracy of about 1\%.

\begin{figure}[t!b]
	\centering
	\vspace{-0.3cm} 
	\subfigtopskip=1pt 
	\subfigbottomskip=-2pt 
	\subfigcapskip=-12pt 
	\subfigure[]{
		\label{structure6 (a)}
		\includegraphics[width=\linewidth]{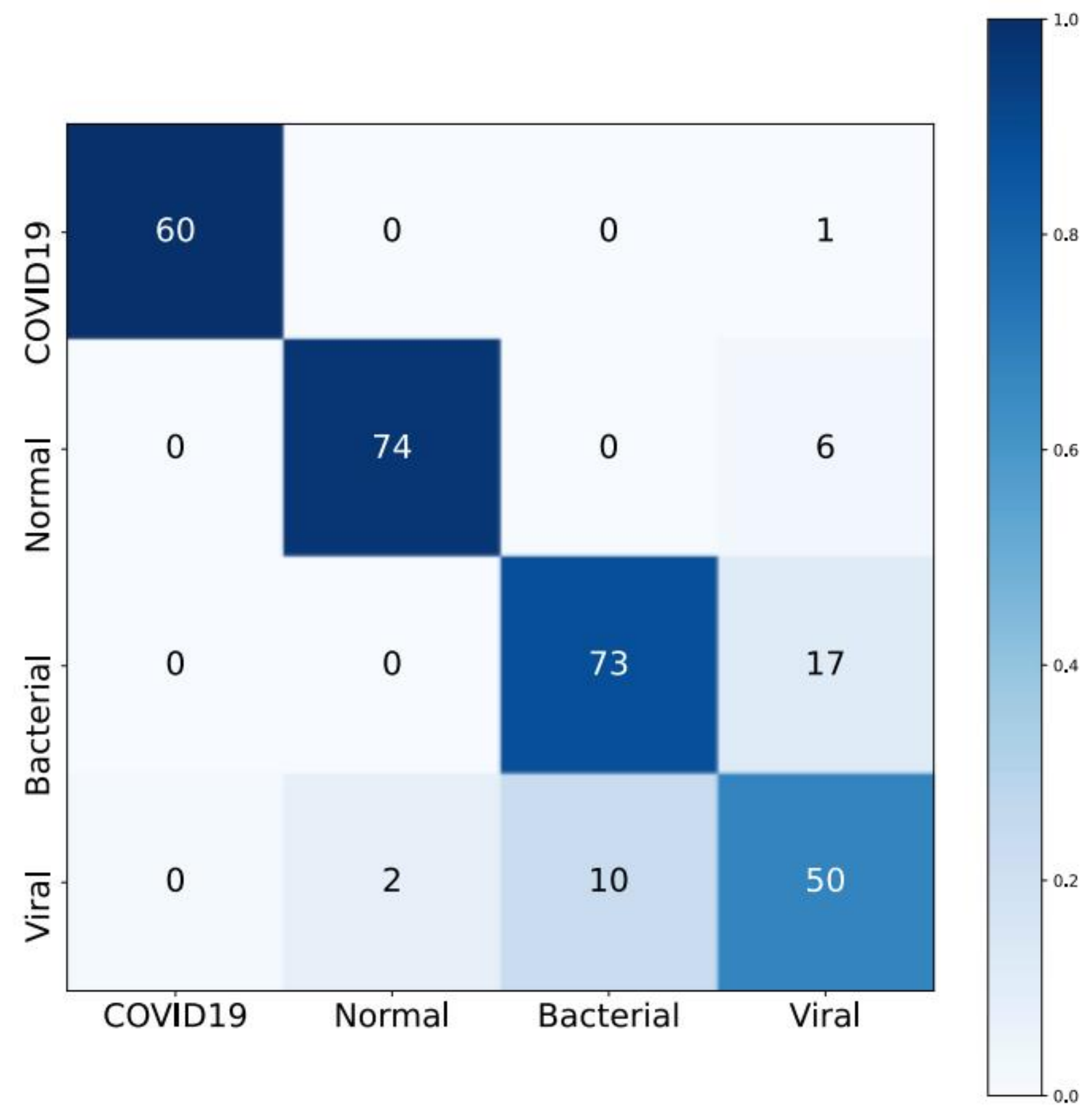}}
	
	\subfigure[]{
		\label{structure6 (b)}
		\includegraphics[width=\linewidth]{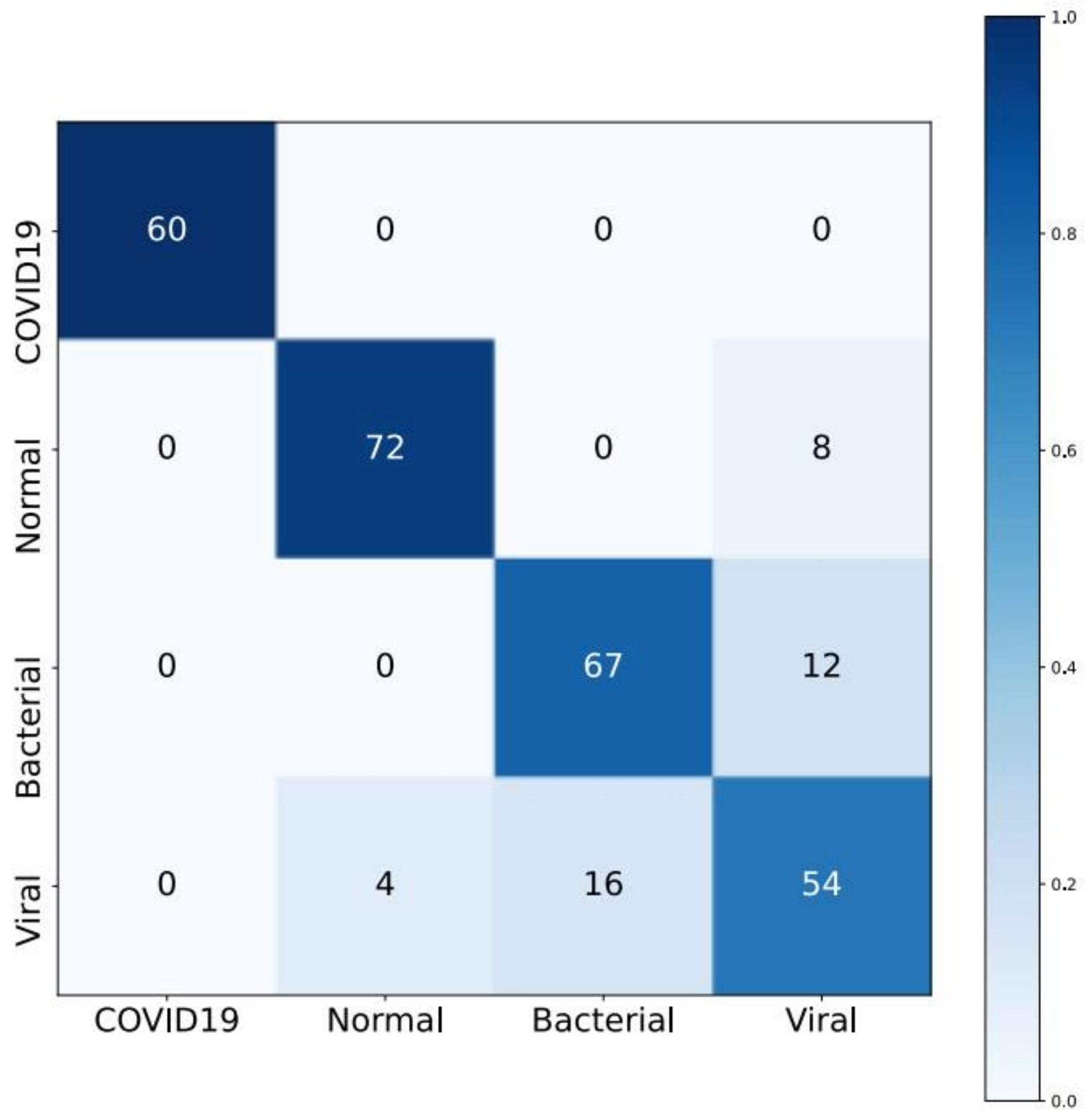}}
	\label{fig:training6_targets}
	\caption{ (a) the results of DPNNet network model training (b) the results of DPNN-SE network model training.}
	\label{the_fig6}
\end{figure}

\begin{table*}[!tbh]
	\centering
	
	\begin{tabular}{|c|c|c|c|c|c|c|c|c|}
		\hline
		\multirow{2}*{Model} & \multicolumn{4}{c|}{Data Enhancement} & \multicolumn{4}{c|}{Without Data Enhancement}\\
		\cline{2-9}
		{}& Accuracy & Precision & Recall & F1-score& Accuracy & Precision & Recall & F1-score\\
		\Xhline{3\arrayrulewidth}
		Resnet18& 0.77 & 0.96 & 0.80 & 0.87& 0.78 & 0.90 & 0.93 & 0.92\\
		\hline
		Resnet50& 0.79 & 0.92 & 0.98 & 0.95& 0.80 & 0.97 & 0.98 & 0.98\\
		\hline
		Resnet101& 0.81 & 0.94 & 0.97 & 0.95& 0.78 & 0.98 & 0.90 & 0.94\\
		\hline
		DenseNet121& 0.82 & 0.98 & 0.98 & 0.98& 0.84 & 0.97 & 0.98 & 0.98\\
		\hline
		DenseNet201& 0.87 & 0.98 & 1.00 & 0.99& 0.86 & 1.00 & 1.00 & 1.00\\
		\hline
		DPN68&0.74&0.97&0.95&0.96&0.81&0.93&0.95&0.94\\
		\hline
		DPN92&0.80&0.89&0.98&0.94&0.80&0.94&1.00&0.97\\
		\hline
		DPN131&0.80&0.92&0.97&0.94&0.83&0.97&0.98&0.98\\
		\hline
		VGG16&0.79&0.91&0.95&0.93&0.81&0.93&0.95&0.94\\
		\hline
		InceptionV3&0.82&0.94&0.97&0.95&0.82&0.94&0.97&0.95\\
		\hline
	\end{tabular}
	\caption{ Use a variety of different network models to test the effects of data enhancement and unused 
		data enhancement. The evaluation indicators include accuracy, precision, recall, F value (precision rate, 
		recall rate, F value and other indicators). -19 categories are regarded as positive examples, others are 
		false examples)}
	\label{table1}
\end{table*}

Our purpose is to improve the recognition accuracy of the network. In addition to 
using data enhancement to increase the recognition accuracy, we also want to modify 
the network model to increase the recognition accuracy. The main framework of the 
network is DPN network [29]. We use a DPN-92 network, which starts with a 7 * 7 
convolution layer and a maximum pooling layer, and then there are four stages, each stage is composed of multiple substages. On the basis of DPN-92, attention mechanism 
network is introduced. The data set during training is processed by the data 
enhancement described in Section 3.1

\begin{figure}[ht!]
	\centering
	\vspace{0.5cm} 
	\subfigtopskip=1pt 
	\subfigbottomskip=3pt 
	\subfigcapskip=-1pt 
	\subfigure[]{
		\label{structure7 (a)}
		\includegraphics[width=0.75\linewidth]{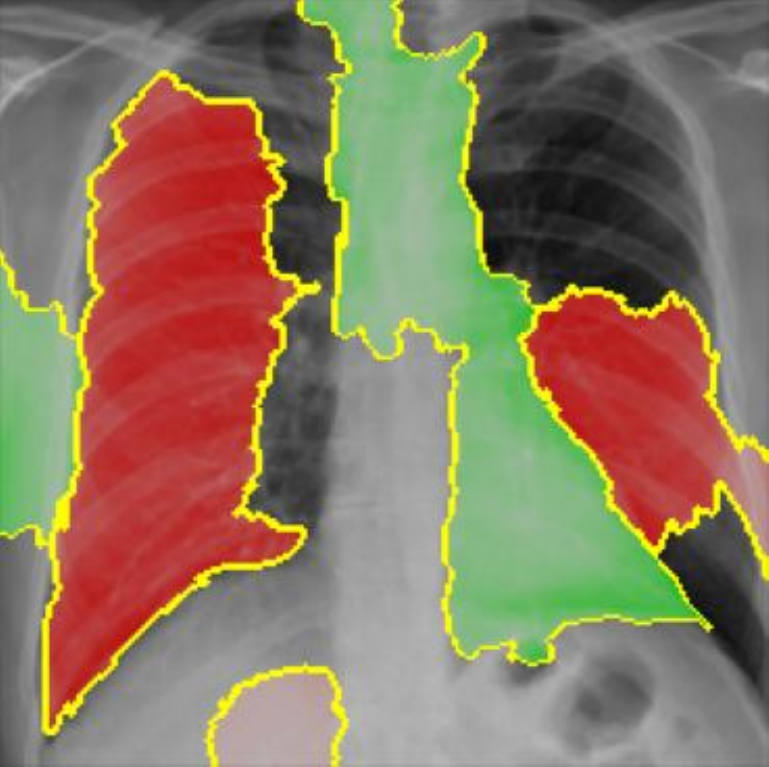}}
	
	\subfigure[]{
		\label{structure7 (b)}
		\includegraphics[width=0.75\linewidth]{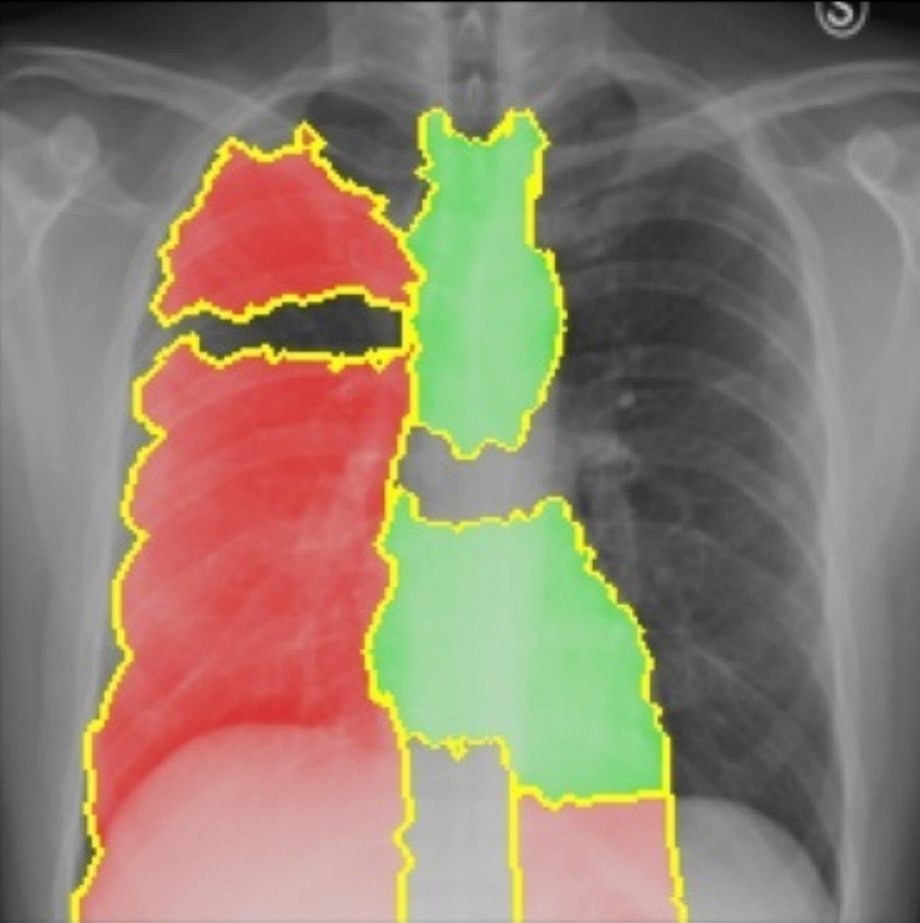}}
	\label{fig:training7_targets}
	\caption{ Examples of COVID-19 model activation maps.}
	\label{the_fig7}
\end{figure}

\begin{table}[!tbh]
	\centering
	
	\begin{tabular}{|c|c|c|c|}
		\hline
		Class&Precision&Recall&F-measure\\
		\Xhline{3\arrayrulewidth}
		COVID-19&0.98&0.98&0.98\\
		\hline
		Normal&0.92&0.95&0.94\\
		\hline
		Pneumonia Bacterial&0.1&0.87&0.78\\
		\hline
		Pneumonia Viral&0.72&0.51&0.60\\
		\hline
		Overall accuracy&\multicolumn{3}{c|}{0.82}\\
		\hline
	\end{tabular}
	\caption{Performance of DPNNet}
	\label{table2}
\end{table}
\begin{table}[!tbh]
	\centering
	
	\begin{tabular}{|c|c|c|c|}
		\hline
		Class&Precision&Recall&F-measure\\
		\Xhline{3\arrayrulewidth}
		COVID-19&0.97&0.98&0.98\\
		\hline
		Normal&.94&0.96&0.95\\
		\hline
		Pneumonia Bacterial&0.1&0.87&0.78\\
		\hline
		Pneumonia Viral&0.74&0.88&0.81\\
		\hline
		Overall accuracy&\multicolumn{3}{c|}{0.84}\\
		\hline
	\end{tabular}
	\caption{Performance of DPN-SE}
	\label{table3}
\end{table}

As shown in Fig.~\ref{the_fig6}, we show the confusion matrix of the recognition effect of 
the classic DPN network and the modified DPN-SE network with the self-attention mechanism added. Table ~\ref{table2} shows the accuracy, precision, recall rate and F-measure of 
DPN on the test set. Table~\ref{table3} shows the accuracy, precision, recall rate and F-measure 
of DPN-SE on the test set. In the confusion matrix, it can be seen that this test set has 
64 cases of new coronary pneumonia, 89 normal cases, 76 cases of viral pneumonia, 
and 64 cases of viral pneumonia. The classification performance of viral pneumonia 
and bacterial pneumonia is lower than that of the other two categories, resulting in a 
lower overall accuracy rate. If we combine bacterial pneumonia and viral pneumonia 
into pneumonia, the overall accuracy will be significantly improved. The good news is 
that our detection of new crowns and normal lungs has a high accuracy rate. The 
average accuracy rate on the DPN network is 82\%, the precision is 98\%, the recall rate 
is 98\%, and the F-measure is 98\%. On the DPN-SE network, the average accuracy 
rate is 84\%, the precision is 97\%, the recall rate is 98\%, and the F-measure is 98\%. The average accuracy rate has improved by 2\%. The positive results obtained in the 
experimental data are about the high accuracy and recall rate of the COVID-19 category. 
A higher recall rate means a lower false negative (FN), and a lower number of false 
negatives (FN) is the result we hope to get. This is very important because minimizing 
missed cases of COVID-19 is an important goal of medical diagnosis. In general, when 
the attention mechanism structure is added to the network model, it can be observed 
that the average accuracy of the DPN-SE network has increased by 2\%, indicating that 
the modified network model has played a role.

We need to clearly mark the features judged by the model on the chest X-ray image, 
so that the doctor can quickly find the lesion area and judge whether the marked part is the basis for judgment. Fig.~\ref{the_fig7} shows an example of activation diagram that can explain 
model processing using lime. This method does not go deep into the model. It detects 
changes in the output of the black box model by slightly perturbing the input, and trains 
an interpretable model at the point of interest (original input) based on this change. 
Only x-ray images and the model that has been trained can get the activation map. The 
decision of the Lime model is represented by a red/blue background, in which the red 
area represents the key feature that the classification model focuses on, and the blue 
area is the area of unnecessary attention. It can be observed from the figure that the red 
key feature areas are basically distributed inside the chest cavity, which requires special 
attention. The blue area is the non-pulmonary area in the middle and edge of the body.

The environment we used for the experiment was 1080TiGPU, and it took 3s to 
analyze the category of a picture and visualize the activation map on a single GPU.

\section{Conclusion}

With the COVID-19 pandemic, the number of cases is increasing. Many places are facing the challenge of shortage of testing resources. In this article, we propose a deep neural network model DPN-SE that uses chest X-rays to identify COVID-19 cases. And for a small number of samples, data enhancement methods are used, and good results have been achieved on the test set. Compared with DPN-SE, DPN-SE has the same computational overhead as the DPN network structure, but it improves the recognition accuracy by about 2\%. When more training samples are used, the performance can be further improved. The accuracy of the recognition results of our model is very high, which is believed to help radiologists have a deeper understanding of case related to COVID-19.

\section*{Source code and dataset}
In order for everyone to continue the research well, we provide the code and data set for the experimental research. The trained model and data can be obtained here:

\href{https://github.com/ChengBo5/covid19-X-ray.git}{{ $\tt https://github.com/ChengBo5/covid19-X-ray.git$}}

\section*{Acknowledgment}
This work is financially supported by the Fundamental Research Funds for Central University, Southwest Minzu University ( 2020NYBPY02).

\small
\bibliographystyle{IEEEtran}
\bibliography{sss}

\end{document}